\documentclass[aps,prl,
 amsmath,amssymb,
 twocolumn, groupedaddress,
 showpacs]{revtex4-1}
\usepackage{bm}
\usepackage{graphicx}
\usepackage{gensymb}
\usepackage{amsmath}
\usepackage{amssymb}
\usepackage{color}
\usepackage{ulem}

\begin{document}

\title{Observation of a Dirac nodal line in AlB$_2$}
\author{Daichi Takane,$^1$ Seigo Souma,$^{2,3}$ Kosuke Nakayama,$^1$ Takechika Nakamura,$^1$ Hikaru Oinuma,$^1$ Kentaro Hori,$^1$ Kouji Horiba,$^4$ Hiroshi Kumigashira,$^{4,5}$ Noriaki Kimura,$^1$ Takashi Takahashi,$^{1,2,3}$ and Takafumi Sato$^{1,2}$}

\affiliation{$^1$Department of Physics, Tohoku University, Sendai 980-8578, Japan\\
$^2$Center for Spintronics Research Network, Tohoku University, Sendai 980-8577, Japan\\
$^3$WPI Research Center, Advanced Institute for Materials Research, Tohoku University, Sendai 980-8577, Japan\\
$^4$Institute of Materials Structure Science, High Energy Accelerator Research Organization (KEK), Tsukuba, Ibaraki 305-0801, Japan\\
$^5$Institute of Multidisciplinary Research for Advanced Materials (IMRAM), Tohoku University, Sendai 980-8577, Japan
}

\date{\today}

\begin{abstract}
     We have performed angle-resolved photoemission spectroscopy of AlB$_2$ which is isostructural to high-temperature superconductor MgB$_2$. Using soft-x-ray photons, we accurately determined the three-dimensional bulk band structure and found a highly anisotropic Dirac-cone band at the $K$ point in the bulk hexagonal Brillouin zone. This band disperses downward on approaching the $H$ point while keeping its degeneracy at the Dirac point, producing a characteristic Dirac nodal line along the $KH$ line. We also found that the band structure of AlB$_2$ is regarded as a heavily electron-doped version of MgB$_2$ and is therefore well suited for fully visualizing the predicted Dirac nodal line. The present results suggest that (Al,Mg)B$_2$ system is a promising platform for studying the interplay among Dirac nodal line, carrier doping, and possible topological superconducting properties.
\end{abstract}


\maketitle

Topological semimetals (TSMs) host a novel quantum state of matter distinct from well-known topological insulators with a finite bulk band gap and metallic Dirac-cone surface states. TSMs are characterized by the band crossing between bulk valence band and conduction band at discrete momentum ($k$) point(s) in the three-dimensional (3D) bulk Brillouin zone (BZ), as realized in 3D Dirac semimetals (DSMs) (e.g. Na$_3$Bi and Cd$_3$As$_2$)  \cite{WangPRB2012, WangPRB2013, NeupaneNC2014, LiuScience2014, BorisenkoPRL2014, RothPRBPRL2018} with four-fold degenerated Dirac nodes and Weyl semimetals (WSMs) (e.g. TaAs family)  \cite{XuScience2015, LvPRX2015, YangNP2015, SoumaPRB2016, PositanoADFM2018} possessing pairs of two-fold degenerated Weyl nodes. Such unconventional band crossings create various exotic physical properties such as chiral anomaly, extremely high mobility, and gigantic linear magneto-resistance \cite{ZyuzinPRB2012, LiuPRB2013, SCZhangPRB2013, ChenPRB013, LandsteinerPRB2014, ChernodubPRB2014, LiangNM2015, XiongScience2015, WengJPhys2016}. There is another category of TSMs characterized by the band crossing along a one-dimensional curve in the $k$ space (nodal line), called topological nodal-line semimetals (TNLSMs). It is theoretically proposed that TNLSMs exhibit unique physical properties such as a flat Landau level, long-range Coulomb interaction, and unusual charge polarization and orbital magnetism \cite{RhimPRB2015, Mitchell2015, HuhPRB2016, RamamurthyPRB2017}. In contrast to the point node, the nodal line can have various shapes in $k$ space depending on the details of the band structure and the type of symmetry protecting the nodes, as highlighted by two-fold Weyl nodal ring \cite{BurkovPRB2011, G. Bian PRB(R)2016}, four-fold Dirac ring \cite{HWengPRB2015}, nodal link \cite{WChenRRB(R)2017, ZYanPRB(R)2017}, and nodal net \cite{XFengPRMat2018, XZhangPRB2017, YiPRB2018}.

  A useful approach to realize such exotic nodal structures is to utilize the band inversion, as demonstrated by the discovery of nodal rings in PbTaSe$_2$ \cite{HasanPbTaSe2}, CaAgAs \cite{CaAgAsTakane}, and (Zr/Hf)SiS \cite{AstNC, HasanPRB(R)ZrSiS, TakanePRB(R)}. However, TNLSMs so-far discovered are characterized by the crossing of energy bands originating from heavy elements such as transition-metal $d$ bands, consequently leading to a finite spin-orbit-gap opening which masks the genuine low-energy properties of nodal fermions. To truly realize the nodal-line-related physical properties, it is important to search for materials which have a nodal structure with a relatively weak spin-orbit coupling (SOC).

  In this regard, AlB$_2$ is one of most suitable materials because the energy bands near the Fermi level ($E_{\rm F}$) are composed of only $s$ and $p$ orbitals of light elements (Al and B). As shown in Fig. 1(a), AlB$_2$ is isostructural to well-known high-temperature superconductor MgB$_2$, consisting of alternatively stacking honeycomb B layers and triangular Al lattices. A first-principles band calculation suggests that AlB$_2$ has a linear band crossing (Dirac cone) which arises from the B sub-lattice at the $K$ point in the bulk BZ \cite{MedvedevaPRB2001}, as in the case of graphene. Due to analogy with multilayer graphene, the band degeneracy at the Dirac point is expected to extend along the $k_z$ direction and form a Dirac nodal line (DNL) along the $KH$ high-symmetry line \cite{LobatoPRB2011} (see Fig. 1b). A similar prediction is also made for isostructural MgB$_2$ \cite{JinarXiv2018}. It is therefore urgently required to experimentally establish the DNL in AlB$_2$-type materials and clarify its possible link to the exotic physical properties.
  
  In this Rapid Communication, we report angle-resolved photoemission spectroscopy (ARPES) of AlB$_2$ single crystal. By utilizing bulk-sensitive soft-x-ray photons from synchrotron radiation, we experimentally established the bulk valence band structure in the 3D bulk BZ, and found a highly anisotropic Dirac-cone energy band in the $k_{x}-k_{y}$ plane which disperses along the $k_z$ direction. This band keeps the degeneracy along the $KH$ line, producing a characteristic DNL. We discuss implications of the present results by comparing with first-principles band-structure calculations and other topological nodal-line materials.
   
\begin{figure}
 \includegraphics[width=3.2in]{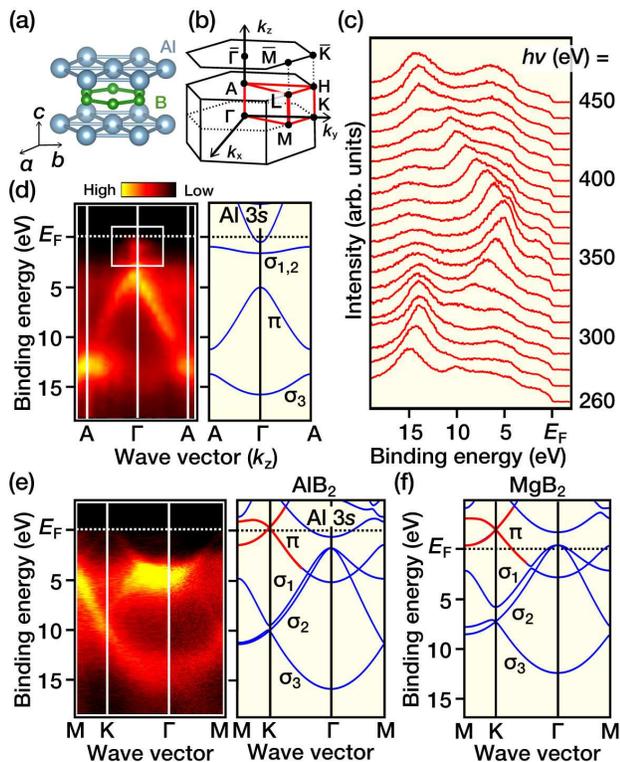}
\caption{(color online). (a) Crystal structure of AlB$_2$. (b) Bulk and (0001) surface BZ of AlB$_2$. (c) Normal-emission EDCs in the valence-band region at $T$ = 40 K measured at the photon energies of 260 - 450 eV. (d) ARPES intensity along the ${\Gamma}A$ line (left), compared with the calculated band structure (right). Area enclosed by white rectangle shows the image around ${\Gamma}$ with enhanced color contrast. (e) ARPES intensity measured along the ${\Gamma}KM$ and ${\Gamma}M$ cuts (left), compared to the calculated band structure of AlB$_2$ (right). Red curve corresponds to the ${\pi}$ band forming the Dirac cone. (f) Calculated band structure of MgB$_2$ along the ${\Gamma}KM$ and ${\Gamma}M$ cuts.}
\end{figure}

High-quality single crystals of AlB$_2$ were grown by the Al-flux method. ARPES measurements were performed with a Scienta-Omicron SES2002 electron analyzer with energy-tunable synchrotron light at BL02 in Photon Factory, KEK. We used linearly polarized light (horizontal polarization) of 260-550 eV. The energy and angular resolutions were set to be 150 meV and 0.2$^\circ$, respectively. Crystals were cleaved in situ in an ultrahigh vacuum better than 1$\times$10$^{-10}$ Torr along the (0001) crystal plane. Sample temperature was kept at $T$ = 40 K during ARPES measurements. The Fermi level ($E_{\rm F}$) of samples was referenced to that of a gold film evaporated onto the sample holder. First-principles band-structure calculations were carried out by QUANTUM ESPRESSO code with generalized gradient approximation \cite{QE}. The plane-wave cutoff energy and the $k$-point mesh were set to be 30 Ry and 24$\times$24$\times$12, respectively.

\begin{figure*}
 \includegraphics[width=6.4in]{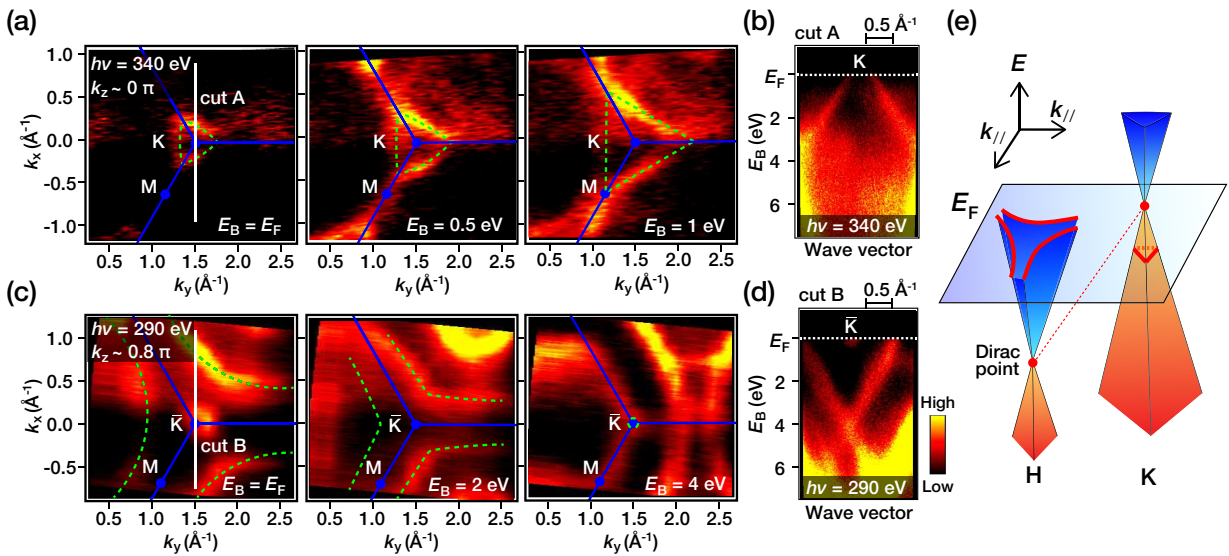}
\caption{(color online). (a) ARPES-intensity mapping as a function of in-plane wave vector ($k_x$ and $k_y$) at selected $E_{\rm B}$ slices, measured at $k_z$ $\sim$ 0 (${\Gamma}KM$ plane) using 340-eV photons. Dashed curve is a guide to the eyes to trace the energy contours of the $\pi$ band. (b) ARPES intensity as a function of wave vector and $E_{\rm B}$, measured along cut A in (a). (c) Same as (a) but measured at $k_z$ $\sim$ 0.8$\pi$ using 290-eV photons. (d) ARPES intensity measured along cut B in (c). (e) Schematic Dirac-cone dispersion in the $k_x$-$k_y$ plane at $k_z$ $\sim$ 0 (right; around the K point) and $\sim$ $\pi$ (left; around the $H$ point).}
\end{figure*}

   We at first performed ARPES measurements along $k$ cut perpendicular to the crystal surface [${\Gamma}A$ cut; see Fig. 1(b)] to elucidate the three-dimensionality of electronic structure. Figure 1(c) shows the energy distribution curves (EDCs) at normal-emission setup measured at $T$ = 40 K with various photon energies ($h\nu$) from 260 to 450 eV. At $h\nu$ = 350 eV, a prominent peak is clearly observed at binding energy ($E_{\rm B}$) of $\sim$ 5 eV, together with a weak hump at around $E_{\rm B}$ $\sim$ 14 eV. The peak rapidly disperses toward higher $E_{\rm B}$ on increasing/decreasing $h\nu$ from 350 eV, while the hump displays a relatively small dispersion. This distinct $h\nu$ dependence of EDCs demonstrates that ARPES measurements with soft-x-ray photons are useful to elucidate the 3D electronic states of AlB$_2$. To better visualize the band dispersions, we plot in Fig. 1(d) the ARPES intensity plotted as a function of $E_{\rm B}$ and $k_z$, compared with the calculated band structure. One can immediately identify a highly dispersive band topped at $\sim$ 5 eV at the ${\Gamma}$ point and a less dispersive band at $\sim$ 14 eV, which are attributed to the B 2$p$ $\pi$ and $\sigma_3$ bands (here we label three $\sigma$ bands as $\sigma_{1}-\sigma_{3}$), respectively, as can be seen from a good agreement in the overall band dispersion between the experiment and calculation. While the calculation predicts weakly dispersive $\sigma_{1,2}$ band at $\sim$ 1 eV, the spectral weight of this band appears to be markedly suppressed in ARPES measurements probably due to the matrix-element effect of photoelectron intensity. A closer look at the ${\Gamma}$ point with enhanced color contrast (area enclosed by white rectangle) signifies the presence of an electronlike band originating from the Al 3$s$ orbital. 

To see the in-plane band dispersions, we plot in Fig. 1(e) the ARPES intensity along the ${\Gamma}KM$ and ${\Gamma}M$ cuts, measured at $h\nu$ = 340 eV (which corresponds to the $k_{z}$ = 0 plane). A reasonable agreement is seen between the experiment and calculation, e.g., in the highly dispersive $\sigma_{3}$ band with its bottom of dispersion at $\sim$ 15 eV at ${\Gamma}$, and the linearly dispersive $\pi$ band near $E_{\rm F}$ around $K$ (highlighted by red curve) which forms the Dirac-cone dispersion, as discussed later in detail. The overall spectral feature is consistent with previous ARPES results with vacuum-ultraviolet (VUV) light on AlB$_2$ and MgB$_2$ \cite{SoumaAlB2, UchiyamaMgB2, SoumaMgB2Nature, TsudaMgB2PRL}. Figure 1(f) shows the calculated band structure for MgB$_2$. One immediately recognizes that the band structure is very similar between AlB$_2$ [right panel of Fig. 1(e)] and MgB$_2$ [Fig. 1(f)] except for the location of $E_{\rm F}$. The $\sigma_{1}$ and $\sigma_{2}$ bands, which are fully occupied in AlB$_2$, are partially occupied in MgB$_2$, and the partially occupied metallic Al 3$s$ band in AlB$_2$ is fully unoccupied in MgB$_2$. This similarity and difference in the electronic band structure between AlB$_2$ and MgB$_2$ is reasonably explained in terms of the same crystal structure with different number of electrons; an Al atom has an extra electron in the outermost shell compared to a Mg atom. The difference in the number of valence electrons provides AlB$_2$ with a great advantage to observe the Dirac-cone dispersion around the $K$ point by ARPES, because the Dirac cone situated near $E_{\rm F}$ in AlB$_2$ is pushed up far above $E_{\rm F}$ in MgB$_2$, becoming unaccessible by ARPES.

To discuss the spectral feature near $E_{\rm F}$ in more detail, we show in Fig. 2(a) the ARPES-intensity mapping as a function of in-plane wave vectors ($k_x$ and $k_y$) at representative $E_{\rm B}$ slices in the $k_{z}\sim 0$ plane (${\Gamma}KM$ plane) measured with 340-eV photons. At $E_{\rm B} = E_{\rm F}$, one can recognize a triangular intensity pattern centered at the $K$ point, which is attributed to the $\pi$ band. This triangular pocket originates from the Dirac-cone band with the Dirac point at $\sim$ 0.5 eV above $E_{\rm F}$ as visible from the band crossing at the $K$ point in Fig. 2(b) [note that the crystal is slightly hole doped since the calculated Dirac point is closer to $E_{\rm F}$, see Fig. 1(e)]. As seen in Fig. 2(a), the small triangular pocket at the $K$ point gradually expands on increasing $E_{\rm B}$ with keeping its triangular shape, reflecting the band-dispersive feature of lower Dirac cone.  This suggests that the Dirac cone is highly anisotropic and reflects the symmetry of the $K$ point ($C_{\rm 3}$). It is noted that a similar triangular-shaped Dirac cone has been observed in doped graphene \cite{McChesneyPRB2010}. Such triangular shape of Fermi surface commonly originates from the $C_{\rm 3}$ symmetry of the $K$ point and the van Hove singularity of the band dispersion at the $M$ point.
 
We have also mapped out the ARPES intensity as a function of in-plane wave vectors at several $k_z$ slices. Figure 2(c) shows the representative energy contours at $k_{z}\sim0.8\pi$ obtained with $h\nu$ = 290 eV. One immediately recognizes that the intensity distribution is very different from that of $k_z$ = 0 plane [Fig. 2(a)] (note that the ARPES intensity at $k_{z}\sim\pi$ was found to be very weak and the energy contour cannot be accurately discussed). This is reasonable since AlB$_2$ has a sizable $k_z$ dispersion [see Fig. 1(d)], and the Dirac point is located well below $E_{\rm F}$ ($E_{\rm B}\sim$ 4 eV) in this $k_z$ slice, as visible from the band dispersion in Fig. 2(d). Since the upper Dirac cone touches another Dirac cone located at the adjacent $\bar K$ point well above the Dirac point, the energy contour at $E_{\rm F}$ no longer shows a closed triangular pattern, but exhibits an open arc-like pattern with no $E_{\rm F}$-crossing of band on the $\bar{K}\bar{M}$  cut. On increasing $E_{\rm B}$, this open contour gradually shrinks toward the $\bar K$point, finally converting into a Dirac point at $E_{\rm B}\sim$ 4 eV (note that there are several other bands that complicate the intensity pattern). Such behavior is schematically illustrated in Fig. 2(e).
 
 \begin{figure*}
 \includegraphics[width=6.4in]{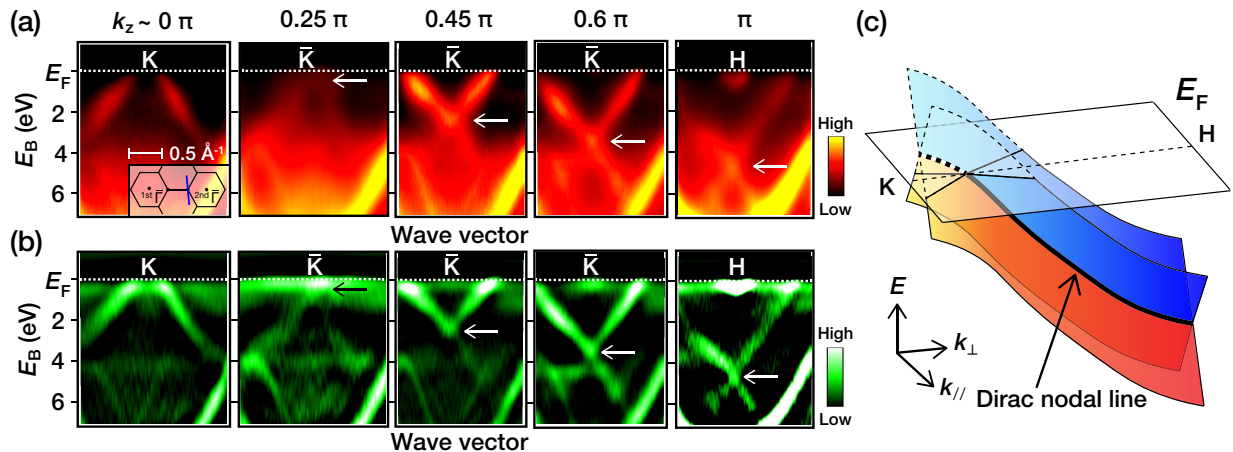}
 \hspace{0.2in}
\caption{(color online). (a) and (b) ARPES intensity and corresponding second-derivative intensity plots, respectively, measured along a cut crossing the $\bar{K}$ point for selected $k_z$'s. Arrows indicate location of the Dirac points. (c) Schematic of the Dirac-cone dispersion and DNL around the $KH$ line.}
\end{figure*}
 
To elucidate how the Dirac-cone band disperses along the $k_z$ direction, we show in Fig. 3(a) the ARPES intensity across the $\bar K$ point measured at various $k_z$'s, together with the corresponding second-derivative intensity plots [Fig. 3(b)]. In Figs. 3(a) and (b), we observe a linearly dispersive band for all the $k_z$ slices. The Dirac point located at $\sim$ 0.5 eV above $E_{\rm F}$ at $k_{z}\sim$ 0 gradually moves downward on increasing $k_{z}$, and finally reaches 5 eV below $E_{\rm F}$ at $k_{z}\sim\pi$. Intriguingly, the band always keeps the degeneracy at the $\bar K$ point irrespective of $k_z$ with no indication of energy-gap opening. This firmly establishes the presence of a DNL along the whole $KH$ line in AlB$_2$. The DNL has a finite slope in the $E$-$k_z$ space due to the $k_z$ dispersion of $\pi$ band, as illustrated in Fig. 3(c). It is noted that we surveyed the electronic states over the entire BZ, and found no evidence for the existence of other DNLs in AlB$_2$. This is reasonable since the Dirac cone exists only around the $\bar K$ point, as recognized from Fig. 1(e).

  Now we discuss characteristics of the observed DNL in AlB$_2$. The observed Dirac-cone band responsible for the DNL originates from the $\pi$ band associated with the honeycomb B layer, which is the same situation as graphene showing the Dirac-cone band with the C-2$p$ $\pi$ character. It was theoretically predicted that when graphene stacks with AA sequence, the Dirac point always keeps the degeneracy irrespective of $k_z$ and forms the DNL along the $KH$ line \cite{LobatoPRB2011} since the sublattice symmetry is preserved in such multilayer graphene. Unfortunately, this situation is not realized in actual graphene since multilayer graphene (graphite) usually stacks with AB or ABC sequence. On the other hand, the honeycomb boron layer in AlB$_2$ stacks with AA sequence, possibly leading to the emergence of DNL in AlB$_2$. However, one may concern that such DNL is easily gapped by the spin-orbit interaction (unless the band degeneracy is protected by nonsymmorphic symmetry), as demonstrated in the gap opening in silicene and germanene \cite{C.-C.LiuPRL2011}. But, in the case of AlB$_2$, the SOC is extremely weak (in the level of $\mu$eV  by referring to the atomic SOC value of boron \cite{JinarXiv2018}), so that such problem is overcome, making AlB$_2$ an ideal material to study the behavior of DNL.

It is inferred that the observed DNL in AlB$_2$ has the same origin as that recently predicted for MgB$_2$ \cite{JinarXiv2018}, because both materials have essentially the same crystal structure and consequently the same band structure except for the position of the chemical potential (Fermi level) due to the difference in the number of valence electrons (Al has one additional electron compared to Mg). It is therefore suggested that MgB$_2$ also has a DNL and may behave as topological superconductor with nodal fermions contributing to the superconducting pairing. It is noted here that AlB$_2$ is a suitable system to investigate the anomalous transport properties originating from non-zero Berry phase, because it is theoretically proposed that the Landau orbit enclosing the $KH$ line gives rise to the non-zero Berry phase \cite{JinarXiv2018}. This condition is exactly satisfied in AlB$_2$ [see Fig. 2(a)] but not in MgB$_2$. Moreover, by tuning the Al/Mg ratio in (Al,Mg)B$_2$ alloy, we are able to systematically control the electronic phase between the superconducting phase and the non-trivial Berry phase without breaking the DNL. The (Al,Mg)B$_2$ system would provide a precious opportunity to investigate the relationship between topological superconductivity and Berry phase.

  Finally, we briefly comment on the relationship between the present results and recent studies on isostructural compounds TiB$_2$ and ZrB$_2$. It has been reported by the first-principles band calculations and ARPES that these compounds have three different types of nodal structures (nodal net) originating from the Ti-3$d$/Zr-4$d$ electrons \cite{XFengPRMat2018, XZhangPRB2017, YiPRB2018} in contrast to the simple DNL in AlB$_2$ originating from the B-2$p$ electrons. While the nodal net of TiB$_2$/ZrB$_2$ is associated with mirror symmetry together with time-reversal and space-inversion symmetry ($PT$ symmetry), the DNL of AlB$_2$ is characterized by the sub-lattice symmetry of honeycomb B layer and $PT$ symmetry. Therefore, the electronic states associated with the nodal fermions and related symmetries are very different between AlB$_2$ and (Ti/Zr)B$_2$, although they share the same crystal structure. The AlB$_2$-type layered diborides would thus serve as an excellent platform to explore a rich variety of nodal fermions and their relationship to symmetries.

In summary, we performed ARPES experiments of AlB$_2$ with bulk-sensitive soft-x-ray photons, and determined the 3D band structure and Fermi surface. We found a triangular-shaped Dirac-cone dispersion at the $K$ point of bulk BZ in the $k_x$-$k_{\rm y}$ plane. The Dirac cone significantly disperses along the $k_z$ direction while keeping the band degeneracy at the $\bar K$ point of surface BZ, producing a characteristic DNL associated with the combination of $PT$ symmetry and sub-lattice symmetry of honeycomb B layer. The present result paves a pathway toward investigating the exotic nodal structures in weakly spin-orbit-coupled topological materials.

\begin{acknowledgements}
This work was supported by Grant-in-Aid for Scientific Research on Innovative Areas ``Topological Materials Science'' (JSPS KAKENHI No: JP15H05853), Grant-in-Aid for Scientific Research (JSPS KAKENHI No: JP17H01139, JP18H01160 and JP18H04472), Grant-in-Aid for JSPS Research Fellow (No: JP18J20058) and KEK-PF (Proposal No: 2018S2-001 and 2016G555). 
\end{acknowledgements}


\bibliographystyle{prsty}

\end{document}